\documentclass[twocolumn,eqsecnum,showpacs,showkeys]{revtex4}
\usepackage[dvips]{graphicx}

\begin{document}
\title{Effect of Interaction on the Formation of Memories in Paste}

\author{Yousuke Matsuo}
\author{Akio Nakahara}
\email{nakahara@phys.ge.cst.nihon-u.ac.jp}
\affiliation{
Laboratory of Physics, College of Science and Technology \\
Nihon University, Funabashi, Chiba 274-8501, Japan
}%

\date{\today}

\begin{abstract}
\ \\
A densely packed colloidal suspension with plasticity,
 called paste,
 is known to remember directions of vibration and flow. 
These memories in paste can be visualized
 by the morphology of desiccation crack patterns. 
Here, we find that paste made of charged colloidal particles
 cannot remember flow direction. 
If we add sodium chloride into such paste
 to screen the Coulombic repulsive interaction between particles,
 the paste comes to remember flow direction. 
That is, one drop of salt water changes memory effect in the paste
 and thereby we can tune the morphology of desiccation crack patterns
 more precisely.
\end{abstract}


\maketitle

\section{Introduction}

Study on crack formation is
 not only an attractive theme for scientists,
 but also very important
 in various fields of technology.
For example,
 when a brittle material such as a glass plate falls to the ground,
 random crack patterns are formed due to rapid impact.
Here, there appear statistical laws such that
 mass distributions of fragments obey scaling laws
\cite{Herrmann90, Lawn93, Oddershede93, Herrmann06}.
When a glass plate is heated and then suddenly cooled,
 contrastively,
 the morphology of crack patterns can be controlled
 as a cooling rate is carefully increased.
A transition from a straight crack to an oscillatory one appears
 at such situations
\cite{Yuse93, Marder94, Hayakawa94, Sasa94}.
It is a wonder that crack phenomena contain such variety of physical laws.

The formation of desiccation cracks of paste also shows
 interesting phenomena.
Here, paste means a densely packed colloidal suspension with plasticity,
 like clay.
It is reported that
 formation of columnar joint is reproduced
 in the drying process of deep starch paste.
That is,
 shrinkages due to cooling and drying induce the same kind of stresses
 which can be released by the formation of similar crack structures
\cite{Muller98, Mizuguchi05, Goehring06}.
Recently, it has been reported that
 paste remembers the direction of external fields
 and the morphology of desiccation cracks
 can be controlled by memory effects of paste
\cite{Nakahara05a, Nakahara06b, Nakahara07}.

If we mix powder with water,
 pour the mixture into a container to make a thin layer,
 and dry it at a room temperature,
 desiccation cracks emerge
 with characteristic sizes of these fragments
 proportional to the depth of the mixture
\cite{Groisman94, Allain95, Komatsu97, Kitsune99}.
When the mixture contains a lot of water,
 it can be regarded as a viscous Newtonian fluid,
 and we usually get isotropic and cellular desiccation crack patterns.
The morphology of crack patterns is sometimes influenced
 by the drying gradient
\cite{Allain95, Shorlin00}.
On the other hand,
 when the mixture contains less water,
 it can be regarded as a non-Newtonian fluid with plasticity,
 so we call it paste.
Due to the plasticity,
 paste remembers the direction of external field,
 such as vibration and flow,
 and the direction of desiccation crack propagation
 is determined by the memory effect of the paste
\cite{Nakahara05a, Nakahara06b, Nakahara07}.

For example,
 a water-poor paste remembers the direction of vibration.
If we vibrate a water-poor paste horizontally
 for some time as short as 1 min.,
 then stop the vibration and dry it slowly at a room temperature
 for a long time from days to weeks,
 the paste is found to remember the direction of the initial vibration,
 so that the desiccation cracks propagate along the direction
 perpendicular to the direction of the initial vibration
 the paste got days or weeks ago.
Thus, we get a regularly orientated lamellar desiccation crack pattern,
 with the directions of lamellar cracks all perpendicular
 to the direction of the initial vibration.
We call this phenomenon "memory effect of paste on vibration".
We confirmed that this memory lasts at least one month
 by starting the drying process one month after the initial vibration
\cite{Nakahara05a}.

What will happen when we vibrate a water-rich paste?
Of course, even if we call this water-rich,
 it contains certain amount of colloidal particles
 as long as it holds plasticity.
In many cases,
 once water-rich paste is fluidized to form a flow pattern
 during the initial vibration,
 the paste cannot remember what happened to it,
 and we get only isotropic and cellular crack patterns.
A paste of calcium carbonate (${\rm CaCO_3}$)
 is a typical example of such pastes.
However,
 some water-rich pastes,
 such as those of magnesium carbonate hydroxide,
 show different phenomena.
When such kind of water-rich pastes are
 fluidized during initial vibration,
 there emerges another type
 of regularly orientated desiccation crack pattern,
 with the direction of crack propagation
 parallel to the direction of the flow induced
 by the initial vibration.
We will demonstrate that
 this kind of water-rich pastes remember the flow direction,
 and call this phenomenon "memory effect of paste on flow".
Since such pastes remember the directions of flow,
 we can imprint (in principle) any flow patterns
 into paste to make various crack patterns,
 such as lamellar, radial, ring, spiral, and so on
\cite{Nakahara06b, Nakahara07}.

Thus, we know that there are two types of memory effects,
 and we need to investigate what is essential to their emergence.
It is reported that
 densely packed inelastic particles exhibit jamming phenomena,
 which can be an origin of irreversible and persistent deformation,
 i.e. plasticity
\cite{Liu98, Liu01, Coussot05, Miguel06}.
As for the memory on vibration of water-poor paste,
 we conjecture that,
 when a water-poor paste is vibrated horizontally,
 a longitudinal density fluctuation emerges
 along the direction of the vibration,
 just like cluster formation of inelastic particles
\cite{Goldhirsh93, Nakahara97}.
Due to the plasticity,
 these anisotropic microstructures remain
 and therefore desiccation cracks run
 along the direction perpendicular to the direction
 of the initial vibration
\cite{Nakahara05a}.
Theoretical and numerical works also confirm that
 plasticity plays an important role in memory effect of paste.
Models which take into account
 plastic deformation of elastic media under external fields
 can explain the memory effect of vibration
\cite{Ooshida05, Otsuki05, Ooshida08}.

Here,
 there is one big question about the other memory effect,
 namely the "memory effect of flow":
 some pastes remember the direction of vibration and flow,
 but others remember only the direction of vibration
 and cannot remember the flow direction.
The emergence of the memory effect of flow
 depends on what kind of colloidal particles we use.
For example,
 as is shown in Table I,
 paste of magnesium carbonate hydroxide has
 memories of vibration and flow,
 while paste of calcium carbonate (${\rm CaCO_3}$)
 has only memory of vibration
 and cannot remember flow direction
\cite{Nakahara06b}.
Until now, there is no theoretical or numerical study
 to explain memory effect of flow.
In this paper,
 we investigate the mechanism of memory of flow experimentally,
 and find out why some paste have no ability
 to remember the flow direction.

\begin{center}
\begin{table}[h]
\caption{Memory effects}
\begin{tabular}{c|c|c} 
 & calcium carbonate & magnesium carbonate \\
 & (${\rm CaCO_3}$) & hydroxide \\ \hline \hline
Vibration & YES & YES \\ \hline
Flow & NO & YES \\ 
\end{tabular}
\end{table} 
\end{center}

\section{Experimental Investigations}

To investigate the reason why paste of ${\rm CaCO_3}$
 cannot remember flow direction,
 we performed series of experiments to find differences
 between the paste of ${\rm CaCO_3}$ and
 that of magnesium carbonate hydroxide
 (Kanto Chemical, Tokyo, Japan).
As is shown in Table I,
 the paste of magnesium carbonate hydroxide can remember flow direction,
 while that of ${\rm CaCO_3}$ cannot.

\subsection{Size distribution of particles}

First, we measured the density of the materials,
 and found that the density of ${\rm CaCO_3}$ is ${\rm 2.72 g/cm^3}$
 and that of magnesium carbonate hydroxide is ${\rm 2.00 g/cm^3}$.
Recently, we found that
 water-rich carbon paste and water-rich kaolin paste have
 also memory of flow\cite{Nakahara11}.
Since the density of carbon is ${\rm 2.0 g/cm^3}$
 and that of kaolin is ${\rm 2.6 g/cm^3}$,
 we consider that the difference in density between two materials
 does not play an important role in the formation of memory of flow.
This idea is supported
 by the experimental result which will be shown in Sec. 2.4.

\begin{figure}
\includegraphics*[width=7cm]{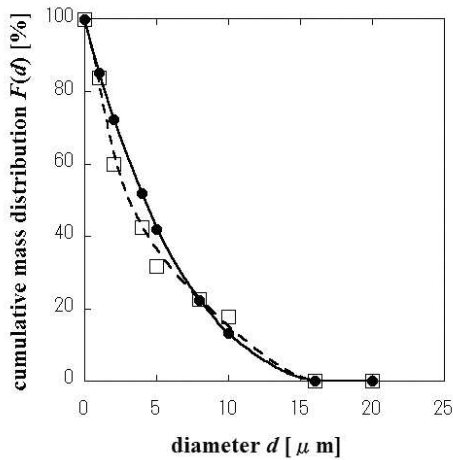}
\caption{
Cumulative mass distribution of colloidal particles, i. e.,
 the fraction of the total mass of colloidal particles
 with diameters larger than $d {\rm [\mu m]}$,
 expressed as $F(d) [\%]$,
 is represented as a function of $d$.
Solid circles and a solid line denote $F(d)$
 of ${\rm CaCO_3}$ particles,
 and open squares and a broken line denotes that of
 magnesium carbonate hydroxide.
}
\label{eps1}
\end{figure}

Next, we measured the size distribution of colloidal particles
 in each paste using sedimentation method based
 on Stokesian approximation.
Figure~\ref{eps1} shows cumulative mass distribution of colloidal particles,
 $F(d)$, i.e., the fraction of total mass of colloidal particles
 with diameters larger than $d {\rm [\mu m]}$. 
Solid circles and the solid line denotes $F(d)$
 of ${\rm CaCO_3}$ particles
 and open squares and the broken line denotes
 that of magnesium carbonate hydroxide.
By taking numerical differentiation of $F(d)$ with respect to $d$,
 we find that both mass distributions of ${\rm CaCO_3}$ particles
 and that of magnesium carbonate hydroxide
 obey Weibull distribution with $k=1$,
 i.e., an exponential distribution.
Here, the mass median diameter,
 defined as the value of $d$ with $F(d) = 50 \%$,
 is $3.8 {\rm \mu m}$ for ${\rm CaCO_3}$ particles
 and $3.3 {\rm \mu m}$ for particles of magnesium carbonate hydroxide,
 respectively.
In addition,
 recently
 we had a chance to purchase two samples of ${\rm CaCO_3}$ pastes
 with different size distributions of colloidal particles.
Size distributions of both ${\rm CaCO_3}$ pastes obey
 exponential distribution,
 but there is a difference in mass median diameter,
 one is $2.0 {\rm \mu m}$ and the other is $4.5 {\rm \mu m}$. 
Since both samples of ${\rm CaCO_3}$ pastes have only memory of vibration
 and do not have a memory of flow,
 we conclude that the difference in the size distribution
 of colloidal particles
 between ${\rm CaCO_3}$ paste and paste of magnesium carbonate hydroxide
 shown in Fig.~\ref{eps1}
 does not play an important role in the formation of memory of flow.

\subsection{Shape of particles}

\begin{figure}
\includegraphics*[width=8.6cm]{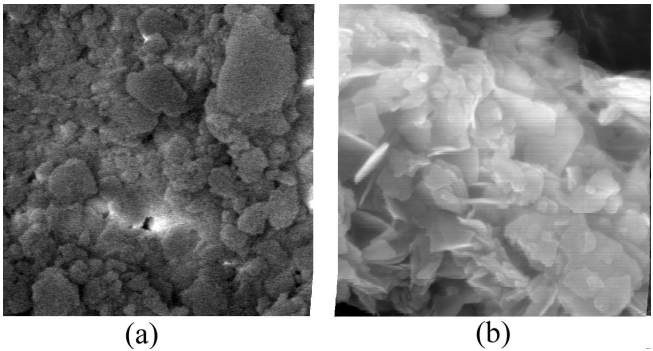}
\caption{
Images taken by Scanning Electron Microscope.
The vertical and horizontal sizes of each figure is 5$\mu$m.
(a) ${\rm CaCO_3}$. (b) Magnesium carbonate hydroxide.
We see that particles of ${\rm CaCO_3}$ look like rough rocks,
 while particles of magnesium carbonate hydroxide
 look like thin plates or disks.
}
\label{eps2}
\end{figure}

We observed the shapes of dry particles of both pastes
 using Scanning Electron Microscope (Hitachi, Tokyo, Japan).
Figure~\ref{eps2} shows that
 particles of ${\rm CaCO_3}$ looks like rough rocks,
 while particles of magnesium carbonate hydroxide
 look like thin plates or disks.
These results suggest us relevance of shape dependence
 on memory effects,
 but we will soon see below that
 the most important factor which determines memory effect
 is not the shape difference.

\subsection{Interaction between particles}

\begin{figure}
\includegraphics*[width=6cm]{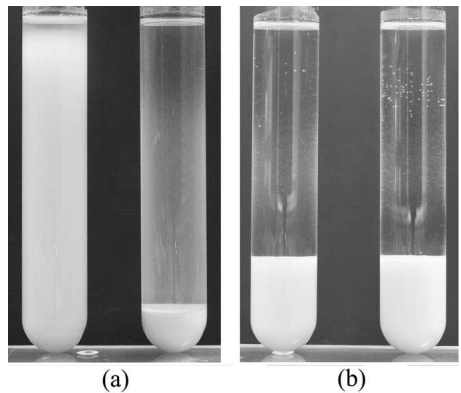}
\caption{
Effect of adding NaCl into dilute suspension
 of colloidal particles and water.
The inner diameter of test glass tube is 18 mm,
 and volume of water in each tube is 80 mL.
Photos are taken 12 hours after we pour the colloidal suspension into tubes.
In (a) and (b),
 left tubes correspond to dilute colloidal suspension
 without NaCl,
 while right tubes correspond to dilute colloidal suspension
 with NaCl.
Here,
 the solid volume fraction of colloidal particles
 in each test tube is ${1.8 \%}$
 and the molar concentration of NaCl in water 
 is given by 0.10 mol/L.
(a) ${\rm CaCO_3}$.
We see no sedimentation of particles in the left tube,
 while in the right tube sedimentation occurs.
These results indicate that particles of ${\rm CaCO_3}$ are charged in water
 so that usually they repel each other
 via the Coulombic repulsive interaction.
Once NaCl is added to the suspension
 like the case of right tube,
 it screens the long-ranged Coulombic repulsive interaction,
 so particles aggregate each other to form clusters that sediment.
(b) Magnesium carbonate hydroxide.
We see sedimentation of colloidal particles in both tubes.
That is,
 particles of magnesium carbonate hydroxide are not charged in water,
 so there is no Coulombic repulsive interaction
 between particles of magnesium carbonate hydroxide in water.
}
\label{eps3}
\end{figure}

We investigated the interaction
 between colloidal particles in water.
Since pastes have plasticity at high solid volume fraction,
 the colloidal particles interact via van der Waals attractive forces
 in both pastes.
Additionally, we found that
 the colloidal particles of ${\rm CaCO_3}$ also interact
 with each other through the Coulombic repulsive interaction.
The particles of ${\rm CaCO_3}$ are charged
 in water\cite{Moulin03, Hoshino08}
 so that usually they repel each other
 via the Coulombic repulsive interaction,
 as shown in the left tube of Fig.~\ref{eps3}(a).
Once we added sodium chloride (NaCl)
 into the colloidal suspension of ${\rm CaCO_3}$,
 it screens the long-ranged Coulombic repulsive interaction,
 so the colloidal particles aggregate each other to form clusters
 that sediment,
 as in the right tube of Fig.~\ref{eps3}(a).
On the other hand, Fig.~\ref{eps3}(b) shows that
 particles of magnesium carbonate hydroxide are not charged in water,
 because there is quick sedimentation of colloidal particles
 of magnesium carbonate hydroxide
 even when we do not add NaCl into the colloidal suspension.

It was already reported that
 particles of ${\rm CaCO_3}$ are charged in water
\cite{Moulin03, Hoshino08}.
Some preliminary works using Zeta potential measurement system
 shows that particles are charged positive in water as for our sample. 
We also applied direct electric field to the paste
 and confirmed that the particles move under electric field.
Recently, we have performed systematic experiments
 using carbon paste
 and found that
 water-rich carbon paste, which is conductive,
 remembers flow direction
\cite{Nakahara11}.
Since water-rich charged pastes cannot remember flow direction,
 while water-rich uncharged pastes or water-rich conductive pastes
 can remember flow direction,
 we realize that the Coulombic repulsive interaction
 prevents the paste to remember flow direction.

\subsection{To remember flow direction}

To observe the effect of screening the Coulombic repulsive interaction
 between the charged colloidal particles,
 we performed drying experiments
 adding NaCl into paste of ${\rm CaCO_3}$.
The mass of powder in paste was fixed at 900 g in each container.
We poured the paste into an acrylic square container
 with 300 mm each side,
 vibrated the container horizontally for 60 s
 at the amplitude of $r = 40 {\rm mm}$ and at a frequency of $f$ [Hz],
 stoped the vibration and dried it
 at a fixed temperature of 25${\rm {}^o C}$ and a humidity of 30\%.

\begin{figure}
\includegraphics*[width=7cm]{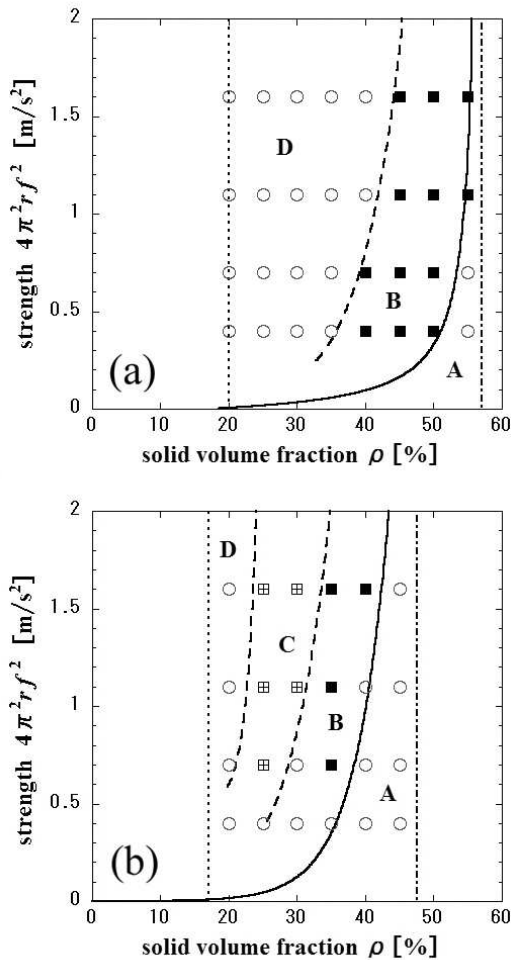}
\caption{
Morphological phase diagrams of desiccation crack patterns,
 shown as functions of the solid volume fraction $\rho$
 and the strength $4 \pi^2 r f^2$ of the initial vibration.
(a) Paste of ${\rm CaCO_3}$.
(b) Paste of ${\rm CaCO_3}$ and NaCl,
 where the molar concentration of NaCl in water is fixed
 at 0.1 mol/L.
Open circles represent isotropic and cellular crack patterns,
 solid squares represent lamellar crack patterns,
 the direction of which is perpendicular
 to the direction of the initial vibration,
 and open squares with a plus inside represent lamellar crack patterns,
 the direction of which is parallel
 to the direction of flow induced by the initial vibration.
The vertical dotted line denotes the Liquid-Limit of the paste,
 and the vertical dash-and-dotted line denotes the Plastic-Limit.
The solid curve represents the strength of initial vibration
 which becomes equal to the value of the yield stress of the paste.
Here, the yield stress of the paste is measured
 by rheometer Physica MCR301 (Anton Paar, Graz, Austria).
In region A below the solid yield stress curve,
 paste did not deform at all
 and so there emerge only isotropic and cellular crack patterns.
The broken curves represent boundaries between regions B, C, and D,
 where pastes remember the directions of vibration in region B,
 pastes remember the direction of flow in region C,
 and we get only isotropic and cellular crack patterns in region D.
Comparing (a) and (b),
 we see that paste of ${\rm CaCO_3}$ cannot remember flow direction,
 but by adding NaCl into paste to screen
 the Coulombic repulsive interaction,
 paste of ${\rm CaCO_3}$ gets an ability to remember flow direction.
}
\label{eps4}
\end{figure}

Figure~\ref{eps4} shows morphological phase diagrams
 of desiccation crack patterns of ${\rm CaCO_3}$ pastes,
 shown as functions of the solid volume fraction $\rho$
 and the strength $4 \pi^2 r f^2$ of the initial vibration.
In Fig.~\ref{eps4}(a) we did not add NaCl into the paste,
 while in Fig.~\ref{eps4}(b) we added NaCl into the paste
 with the molar concentration of NaCl in water
 fixed at 0.1 mol/L.

The vertical dotted line in Fig.~\ref{eps4}
 denotes the Liquid-Limit (LL) of paste,
 and the vertical dash-and-dotted line denotes the Plastic-Limit (PL).
We notice that the values of LL and PL decrease
 as we add more NaCl to the paste of ${\rm CaCO_3}$.
This result means that
 ${{\rm Cl}^-}$ ions screen the Coulombic repulsion interaction
 between the positively charged ${\rm CaCO_3}$ particles
 so that the paste can form network structure
 at lower solid volume fraction.

The solid curve represents the strength of initial vibration
 which becomes equal to the value of the yield stress of the paste.
Here, the yield stress of the paste was measured
 as a function of the solid volume fraction $\rho$ of paste
 by using rheometer Physica MCR301 (Anton Paar, Graz, Austria).
Thus, in region A below the solid yield stress curve,
 the strength of the applied shear stress is smaller
 than that of the yield stress of the paste,
 so the pastes did not deform at all during the initial vibration,
 and we get only isotropic and cellular crack patterns.

The broken curves represent boundaries between regions B, C, and D,
 where pastes remember the directions of vibration in region B,
 pastes remember the direction of flow in region C,
 and we get only isotropic and cellular crack patterns in region D.
These broken curves are drawn on the basis of the data
 of desiccation crack patterns
 to guide eye,
 but once we visualize the motion of paste using black carbon powder
 as will be shown in Sec. 3.2,
 we will find that the boundary between region B and region C (or D)
 corresponds to the threshold
 where the motion of the paste changes from vibration to flow. 

In region B where the value of the applied shear stress is
 just above that of the yield stress of the paste
 and the paste is vibrated just like an earthquake,
 the paste gets a memory of vibration,
 which will be visualized
 as regularly orientated lamellar desiccation crack patterns,
 all perpendicular to the direction of the initial vibration.
 
When we vibrated the paste more hardly
 or vibrated water-richer paste,
 the paste is fluidized to form a flow pattern
 during the initial vibration.
Data in region D of Fig.~\ref{eps4}(a) shows that,
 once ${\rm CaCO_3}$ paste is fluidized, 
 the paste cannot remember what happened to it,
 and only isotropic and cellular crack patterns appear
 in the drying process.
Note that, in region D of Fig.~\ref{eps4}(a),
 we see two types of flow motion:
 one is the one-dimensional flow along the direction of vibration
 and the other is the turbulent flow,
 but in both cases the paste cannot remember anything,
 and we get the same kinds of isotropic and cellular crack patterns.
On the other hand,
 when we added NaCl into paste of ${\rm CaCO_3}$,
 we got different results,
 depending on which kind of flow motion the paste experienced
 at the initial vibration.
In region C of Fig.~\ref{eps4}(b)
 where water-rich paste is fluidized
 and the one-dimensional flow emerges,
 paste remembers the direction of the one-dimensional flow,
 and a regularly orientated lamellar desiccation crack pattern emerges,
 with the crack propagation parallel
 to the direction of the flow.
In region D of Fig.~\ref{eps4}(b)
 where water-rich paste is fluidized
 and the turbulent flow emerges,
 we get only random cellular desiccation crack patterns.
This crack pattern might be a visualization
 of the memory of the turbulent flow,
 but it is difficult to establish a relation
 between the turbulent flow and the random cellular crack pattern.

\begin{figure}
\includegraphics*[width=8.6cm]{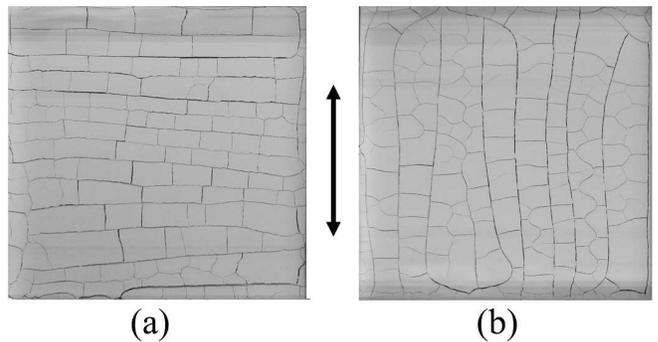}
\caption{
Desiccation crack patterns of paste of ${\rm CaCO_3}$ and NaCl,
 where the molar concentration of NaCl in water is fixed
 at 0.1 mol/L.
The arrow between (a) and (b) indicates the direction of the initial vibration,
 where the amplitude $r$ and the frequency $f$ of the vibration are 40 mm and 1 Hz,
 respectively, i.e., the strength of the initial vibration
 $4 \pi^2 r f^2$ is 1.6${\rm m/s^2}$.
Sides of both square containers are 300 mm in length.
 (a) The solid volume fraction $\rho$ = 35\%.
The direction of lamellar cracks is perpendicular
 to the direction of the initial vibration.
That is, the paste remembers the direction of the initial vibration.
 (b) The solid volume fraction $\rho$ = 30\%.
The direction of lamellar cracks is parallel
 to the direction of the flow induced by the initial vibration.
That is, the paste remembers the flow direction.
These results show that,
 by adding NaCl into paste of ${\rm CaCO_3}$
 to screen the Coulombic repulsive interaction,
 the paste can remember
 not only the direction of the vibration
 but also the flow direction.
}
\label{eps5}
\end{figure}

The distinction between the two memory effects in the regions B and C
 in Fig.~\ref{eps4}(b) is shown in Fig.~\ref{eps5}
 as clearly different desiccation crack patterns.
In Fig.~\ref{eps5}(a) the cracks are perpendicular
 to the direction of the initial vibration,
 while the cracks in Fig.~\ref{eps5}(b) are parallel
 to the flow direction induced by the initial vibration.

To summarize this subsection,
 paste of ${\rm CaCO_3}$ cannot remember flow direction,
 but by adding NaCl into paste of ${\rm CaCO_3}$
 to screen the Coulombic repulsive interaction
 between charged colloidal particles,
 the paste gets an ability to remember flow direction.

\subsection{Critical NaCl concentration for memory of flow}

Is there any critical concentration of sodium chloride (NaCl)
 at which the qualitative behavior of the system changes?
To answer this question,
 we performed systematic experiments
 changing both the molar concentration of NaCl
 and the solid volume fraction of ${\rm CaCO_3}$ particles in paste.
Here,  the amplitude $r$ and the frequency $f$ of the initial vibration
 are 40 mm and 50 rpm (0.83 Hz), respectively,
 and thus the strength $4 \pi^2 r f^2$ of the initial vibration is
 given by 1.1 ${\rm m/s^2}$.

\begin{figure}
\includegraphics*[width=7.5cm]{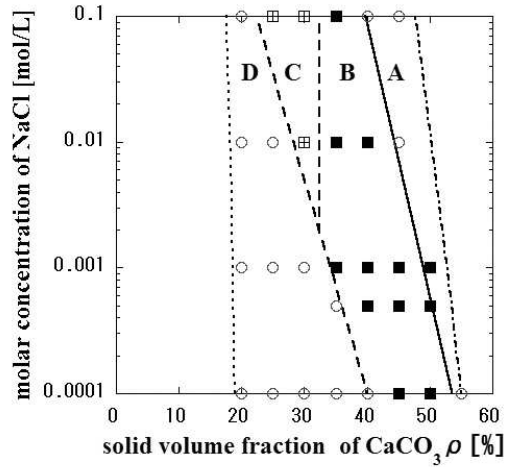}
\caption{
Morphological phase diagram of desiccation crack patterns,
 shown as a function of the solid volume fraction $\rho$
 of ${\rm CaCO_3}$ particles
 and the molar concentration of sodium chloride (NaCl) in water.
Here, the amplitude $r$ and the frequency $f$ of the initial vibration
 are 40 mm and 50 rpm (0.83 Hz), respectively,
 and thus the strength $4 \pi^2 r f^2$ of the initial vibration is
 given by 1.1 ${\rm m/s^2}$.
Open circles represent isotropic and cellular crack patterns,
 solid squares represent lamellar crack patterns,
 the direction of which is perpendicular
 to the direction of the initial vibration,
 and open squares with a plus inside represent lamellar crack patterns,
 the direction of which is parallel
 to the direction of flow induced by the initial vibration.
The almost-vertical dotted line near $\rho = 20 \%$
 denotes the Liquid-Limit (LL) of the paste,
 and the dash-and-dotted line near $\rho = 50 \%$
 denotes the Plastic-Limit (PL) of the paste.
The solid line represents the yield stress line,
 on which the strength of initial vibration
 becomes equal to the value of the yield stress of the paste.
In region A between the solid yield stress line
 and the dash-and-dotted PL line,
 paste did not deform at all
 and so there emerge only isotropic and cellular crack patterns.
The broken curves represent boundaries between regions B, C, and D,
 where pastes remember the directions of vibration in region B,
 pastes remember the direction of flow in region C,
 and we get only isotropic and cellular crack patterns in region D
 due to the emergence of turbulent flows.
Note that, when the molar concentration of NaCl is 0.001 mol/L or below,
 paste of ${\rm CaCO_3}$ has no memory of flow.
By increasing the molar NaCl concentration
 to screen the Coulombic repulsive interaction,
 ${\rm CaCO_3}$ paste gets an ability to remember flow direction
 when the molar concentration of NaCl reaches 0.01 mol/L
 or exceeds that value.
There is a threshold of molar NaCl concentration
 between 0.001 and 0.01 mol/L,
 above which ${\rm CaCO_3}$ paste gets an ability
 to remember flow direction.
}
\label{eps6}
\end{figure}

Figure~\ref{eps6} shows that,
 when the molar concentration of NaCl is 0.001 mol/L or below,
 paste of ${\rm CaCO_3}$ has no memory of flow.
By increasing the molar NaCl concentration
 to screen the Coulombic repulsive interaction,
 ${\rm CaCO_3}$ paste gets an ability to remember flow direction
 when the molar concentration of NaCl reaches 0.01 mol/L
 or exceeds that value.
There is a threshold of molar NaCl concentration
 between 0.001 and 0.01 mol/L,
 above which paste gets an ability to remember flow direction.

\subsection{Summary of experimental results}

We have performed systematic experiments to investigate
 the essential features for the paste to remember flow direction.
We find that interactions between colloidal particles
 plays the dominant role in memory effect.
That is,
 the Coulombic repulsive interaction between colloidal particles
 prevents the paste to remember flow direction.
By adding NaCl into paste
 to suppress the Coulombic repulsive interaction,
 the water-rich paste gets an ability to remember flow direction.

\section{Mechanism of the Memory Effects}

Here, let us discuss the mechanism of memory effects,
 especially the mechanism how water-rich paste
 without the Coulombic repulsive interaction
 can remember flow direction.

\subsection{Interpretation of the mechanism}

\begin{figure}
\includegraphics*[width=8.6cm]{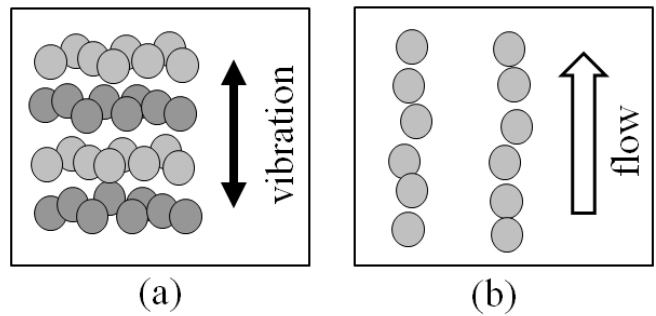}
\caption{
Schematic illustration of microstructures
 which are supposed to keep the memories.
(a) Memory of vibration (b) Memory of flow
}
\label{eps7}
\end{figure}

As for the memory on vibration of water-poor pastes,
 we conjecture that,
 when a water-poor paste is vibrated horizontally,
 a longitudinal density fluctuation emerges
 along the direction of the vibration,
 being similar to stripe formation by collisions
 of crowded inelastic particles
\cite{Goldhirsh93, Nakahara97, Reis02, Sanchez04, Ciamarra07}.
The microscopic structure
 which remembers the direction of vibration
 is illustrated schematically in Fig.~\ref{eps7}(a).
Due to the plasticity,
 this anisotropic microstructure remains in the paste
 and therefore desiccation cracks run
 along the direction perpendicular
 to the direction of the initial vibration.

On the other hand,
 as for the memory on flow of water-rich pastes
 without the Coulombic repulsive interaction,
 we suppose that
 an elongation of a dilute network of colloidal particles
 is the origin of memory of flow,
 as is illustrated in Fig.~\ref{eps7}(b).
When the colloidal particles are charged in water
 and repel each other via the Coulombic repulsive interaction,
 they cannot form such a dilute network structure,
 as is suggested by the experiments in Sec. 2.
When particles are not charged in water,
 particles which attract each other via van der Waals forces
 can form a dilute network structure, such as chains
\cite{deGennes70, Tanaka00, Butter03}.
This dilute network structure can be elongated along flow direction,
 which can be visualized as lamellar desiccation crack patterns,
 the direction of which is parallel to the flow direction.

\subsection{Distinction between memories of vibration and flow}

To establish a clear distinction between the memories
 of "vibration" and "flow",
 we visualize movements of paste under vibration or flow
 by the deformation of the shape of letters
 written on the surface of the paste.
As a paste which remembers directions of both vibration and flow,
 we use paste of magnesium carbonate hydroxide,
 whose morphological phase diagram of desiccation crack patterns
 was presented in Ref. 13
 and is similar to the one shown in Fig.~\ref{eps4}(b).
To write a letter on white paste, we use black carbon powder.

\begin{figure}
\includegraphics*[width=8.6cm]{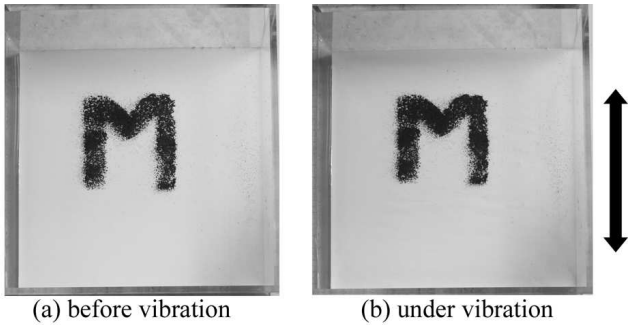}
\caption{
Visualization of the vibrated motion of paste of magnesium carbonate hydroxide
 by the deformation of the letter M
 written by black carbon powder.
The arrow indicates the direction of the external vibration,
 where the amplitude $r$ and the frequency $f$ of the vibration are 15 mm and 2 Hz,
 respectively, i.e., the strength $4 \pi^2 r f^2$ is 2.4${\rm m/s^2}$.
The mass of powder in each container is fixed as 100g,
 and the lengths of the sides of both square containers
 are 200 mm.
The solid volume fraction $\rho$ of the paste is 12.5\%.
At this situation,
 paste of magnesium carbonate hydroxide remembers the direction of the vibration
\cite{Nakahara06b}.
 (a) Before vibration.
 (b) Under vibration, at time $t =$1 min.
 after the onset of the external vibration.
When paste memorizes the direction of the vibration,
 we see only weak deformation of the letter M,
 indicating that the value of the strain is small
 as long as the paste remembers the vibration.
}
\label{eps8}
\end{figure}

Figure~\ref{eps8} shows the visualization
 of the vibrated motion of a water-poor paste.
When the water-poor paste is vibrated
 and the memory of the vibration is imprinted into the paste,
 we see only weak deformation of the letter M,
 indicating that the value of the strain
 in the horizontal plane is small
 when the paste remembers the vibration.
This suggests that
 the memory of vibration remains
 in the form of a small configurational change
 of colloidal particles at microscopic level,
 which is related to a small plastic deformation.

\begin{figure}
\includegraphics*[width=8.6cm]{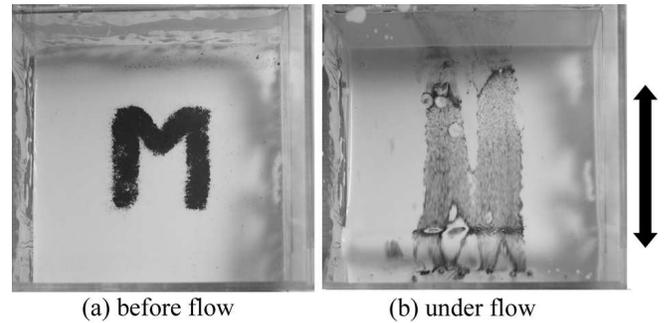}
\caption{
Visualization of the flow motion of paste of magnesium carbonate hydroxide
 by the deformation of the letter M
 written by black carbon powder.
The arrow indicates the direction of the external vibration,
 where the amplitude $r$ and the frequency $f$ of the vibration are
 15 mm and 2 Hz,
 respectively, i.e., the strength $4 \pi^2 r f^2$ is 2.4${\rm m/s^2}$.
The mass of powder in each container is fixed as 100g,
 and the lengths of the sides of both square containers
 are 200 mm.
The solid volume fraction $\rho$ of the paste is 6.7\%.
At this situation,
 paste of magnesium carbonate hydroxide remembers the direction of flow
 induced by the external vibration
\cite{Nakahara06b}.
 (a) Before flow.
 (b) Under flow, at time 3 sec.
 after the onset of the external vibration.
We see that the letter M is elongated along the flow direction.
This means that,
 when paste memorizes the direction of its flow motion,
 there is large scale elongation of the paste
 along the direction of the flow induced by the external vibration.
}
\label{eps9}
\end{figure}

Next, Fig.~\ref{eps9} shows the visualization
 of the flow motion of a water-rich paste.
We see that the letter M written
 on the surface of the water-rich paste
 is elongated along the direction of the flow
 induced by the external vibration.
This means that,
 when the water-rich paste memorizes the direction of its flow motion,
 there is large scale elongation of the paste.
We consider that
 such large scale elongation induces
 an elongation of the network structure
 at mesoscopic level, too.
These macroscopic visualizations of the movements of paste,
 shown in Figs.~\ref{eps8} and ~\ref{eps9}
 will be a guide to distinguish "memory of flow"
 from "memory of vibration",
 and will support our interpretation schematically
 illustrated in Fig.~\ref{eps7}.

\subsection{Pure flow experiments}

There comes a question on
 how we can distinguish "the memory of flow"
 from "the memory of vibration"
 as long as the flow is induced by vibrating the container. 
To distinguish "memory of flow" from "memory of vibration",
 we had once performed an experiment
 where we only applied flow motion to the paste
 without using vibration\cite{Nakahara06b}.
We set the paste in a circular container,
 put a cover directly on the paste,
 kept the container at rest
 and rotated only the cover counter-clockwise for a short time
 to apply a shear flow to the paste,
 and then removed the cover for drying.
After we dried the paste for one week,
 we got a ring-like desiccation crack pattern,
 which is exactly a visualization of the circular flow motion.
This result shows that
 the paste remembers the direction of the flow,
 and the memory of flow is different from the memory of vibration.

\begin{figure}
\includegraphics*[width=7.5cm]{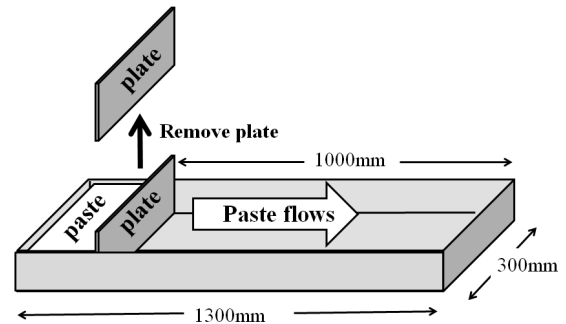}
\caption{
Experimental setup for pure flow experiment.
}
\label{eps10}
\end{figure}

\begin{figure}
\includegraphics*[width=8.6cm]{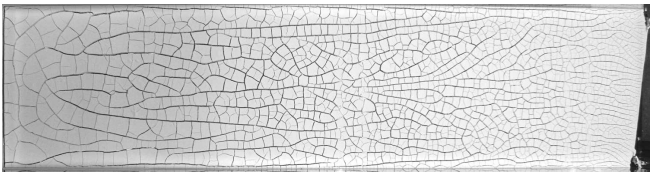}
\caption{
Experimental result of pure flow experiment,
 the method of which was described in Fig.~\ref{eps10}.
Here, we use water-rich paste of magnesium carbonate hydroxide,
 the most typical paste for memory effect of flow.
The paste is prepared
 by mixing 750g of powder with 4500g of distilled water,
 so that the value of the solid volume fraction becomes 7.7$\%$.
At the beginning,
 we stored the paste in a small area located
 left of the vertical plate in Fig.~\ref{eps10}.
After we removed the vertical plate upward rapidly,
 the water-rich paste flows rightward,
 whose movement is just a one-dimensional flow without vibration.
By checking the morphology of the resultant desiccation crack pattern,
 we can confirm that this lamellar crack pattern is
 induced by the "memory of flow",
 and not by the memory of other motions like vibration.
}
\label{eps11}
\end{figure}

There might be still a criticism on the effect of removing the cover
 after the rotation. 
So, we propose another pure flow experiment
 as is shown in Fig.~\ref{eps10}.
At the beginning,
 we stored the paste in a small area located
 left of the vertical plate.
After we removed the vertical plate upward rapidly,
 the water-rich paste flows rightward,
 thus producing a one-dimensional flow without vibration.
Figure~\ref{eps11} shows the morphology
 of the resulting desiccation crack pattern,
 in which the directions of crack propagation are
 parallel to the one-dimensional flow from left to right.
Thus,
 we can confirm that this lamellar crack pattern is
 induced by the "memory of flow",
 and not by the memory of other motions like vibration.

\subsection{Critical NaCl concentration and the Debye length}

Can we relate the value of the critical NaCl concentration
 to some other physical parameters
 to explain the mechanism of the memory of flow quantitatively?
The Debye screening length for NaCl solution is given by
 ${\rm 0.304 / \sqrt{ [NaCl] }}$ nm,
 where [NaCl] is a molar concentration of NaCl in the solution,
 and the values of the Debye length are
 0.96 nm at 0.1 mol/L, 3.0 nm at 0.01 mol/L, 
 9.6 nm at 0.001 mol/L
 and 30.4 nm at $10^{-4}$ mol/L, respectively,
 and 960 nm in totally pure water at pH 7\cite{Israelachvili11}. 
Figure~\ref{eps6} in Sec.2.5 shows that
 the critical NaCl concentration lies between 0.001 and 0.01 mol/L,
 indicating that the corresponding Debye screening length is
 about 5 nm.
The value of 5 nm in the Debye length is
 much smaller than the mass median diameter 3.8 ${\rm \mu m}$
 of the mass distribution function of ${\rm CaCO_3}$ particles.
It is also much smaller
 than the mean interparticle distance 6.0 ${\rm \mu m}$
 for the case of the solid volume fraction $\rho = 30 \%$,
 the typical solid volume fraction for the memory of flow.
Currently,
 it is still difficult to relate the Debye length
 to some characteristic distances in paste.
Probably we must take into account
 the dynamics of the colloidal particles under steady shear flow
 to find a proper non-dimensional parameter
 which describes the transition in the memory effect of flow.

\subsection{Characteristic time for the memory effects}

Are there any characteristic times for the memory effects?
We had confirmed that this memory lasts
 at least one month\cite{Nakahara05a}. 
In that experiment, we set a container in a small box,
 poured a paste into the container,
 vibrated the container for a short time,
 and kept the container at rest for one month. 
Since the container was kept in a small box,
 the humidity inside the small box became soon saturated,
 and the paste was kept wet without drying for one month. 
One month later, we opened the cover of the small box
 to start the drying process.
We got a regularly oriented lamellar desiccation crack patters,
 with the direction of crack propagation perpendicular
 to the direction of the vibration
 that we had added one month before.
This experimental result showed that the memory lasts
 at least one month, and maybe more. 

Then, why does the memory last for such a long time?
We consider that this is due to the plasticity of paste.
The network structure of densely packed colloidal suspension
 does not vanish due to the van der Waals attractive interactions.
Thermal fluctuation cannot destroy the network structure,
 because the sizes of colloidal particles are larger
 than ${\rm 1 \mu m}$.
Rather, the strength of the network increases due to the aging effect,
and this aging phenomena is observed as a thixotropy
by our preliminary rheological measurements of the paste.

\subsection{Future plans}

Microscopic observation of the microstructure inside paste
 is under progress.
We set a microscope below the rheometer Physica MCR301
 to observe structural change of the network structure of paste
 under vibration or shear,
 but it is still difficult to get clear results.
The most difficult factor in this approach is that
 the anisotropy inside the paste is so faint to observe directly.
It is supposed to be a weak density fluctuation
 imposed on a jammed random structure.
Also, as long as paste has plasticity,
 the paste is so cloudy due to the existence of many solid particles
 that it is hard to see inside by light.
Ultra Small Angle Neutron Scattering technique might be needed
 for further investigation.
Currently,
 the formation of desiccation cracks remains the only way
 to visualize the memories in paste.

\section{Concluding Remarks}

We found that the Coulombic repulsive interaction
 between colloidal particles
 prevents paste to remember flow direction.
By adding NaCl into paste
 to suppress the Coulombic repulsive interaction,
 colloidal particles are allowed to form a dilute network structure
 which can be elongated along flow direction,
 and the paste gets an ability to remember the flow direction.
We can now control memory effects in paste
 and tune the morphology of desiccation crack patterns
 locally and more precisely.

The memory effect of paste is going to be important
 in the fields of technology,
 and it is already applied to measure the velocity of desiccation cracks.
Usually, desiccation cracks take cellular and complex structures
 with many branches,
 so it becomes difficult to measure the velocity of these winding cracks
 quantitatively.
When we align the direction of crack propagation by memory effect,
 however,
 we can measure velocities of these straight cracks
 easily and accurately\cite{Kitsune09}.
Other attempts have been challenged
 to control the direction of crack propagation
 by using electrical field\cite{Mal07}
 and magnetic field
\cite{Pauchard08, Ngo08a}.
Combination of these methods enables us to open new stage
 to control morphology of crack patterns.

In the fields of geosciences,
 memory effects of clay pastes will be a useful tool
 to know what happened before in the history of the earth,
 because crack patterns of clay rocks show
 how they were vibrated and fluidized by previous earthquakes.
Even when there appear no cracks,
 we consider that
 clay pastes keep their memories
 in the form of anisotropic microstructure.
If we can get information on memories in clay rocks,
 the history of earthquakes might be estimated.
Since cracks will run according to memories in pastes,
 we might be able to predict
 how rocks and mountains will be destroyed at upcoming huge earthquake
 by studying memories in clay rocks.
We would like to prevent and reduce disasters 
 which will be induced by such diastrophism.

\section*{Acknowledments}

We would like to acknowledge
 Y. Shinohara, K. Hoshino and H. Nakayama
 for performing experiments with us
 and Ooshida Takeshi, S. -I. Sasa, M. Otsuki, S. Kitsunezaki,
 S. Goto, T. Matsumoto, M. K\"ulzer, T. Hatano, N. Ito, Ferenc Kun,
 S. Tarafdar, T. Dutta, M. Yamanaka, Y. Takano, C. Itoi and H. Takahashi
 for valuable discussions.
We also thank
 K. Yoshizawa, T. Okuzono and J. Yamanaka
 for Zeta potential measurements.
This work was supported by Grant-in-Aid for Scientific Research (KAKENHI)
 (B) 22340112 and (C) 21540388 and 23540452
 of Japan Society for the Promotion of Science (JSPS).
This project was also supported by JSPS and HAS
 under the Japan - Hungary Research Cooperative Program
 and by JSPS and DST
 under the Japan - India Science Cooperative Program.

\end{document}